\documentclass[conference]{IEEEtran}
\IEEEoverridecommandlockouts
\usepackage{cite}
\usepackage{textcomp}
\usepackage[utf8]{inputenc}
\usepackage{graphicx}
\usepackage{textcomp}
\usepackage{indentfirst}
\usepackage{algorithmic}
\usepackage{amsmath,amssymb,amsfonts}
\usepackage[ruled,vlined]{algorithm2e}
\usepackage[export]{adjustbox}
\usepackage{xcolor}
\usepackage[left=2cm, right=2cm, top=2cm, bottom=2cm]{geometry}
\def\BibTeX{{\rm B\kern-.05em{\sc i\kern-.025em b}\kern-.08em
    T\kern-.1667em\lower.7ex\hbox{E}\kern-.125emX}}
    
\usepackage{comment}  
\usepackage{dblfloatfix} 
\usepackage{caption}
\usepackage{subcaption}
\usepackage{booktabs}
\usepackage[export]{adjustbox}
\usepackage{epsfig}

\newlength\myindent
\begin{document}

\title{DeSMP: Differential Privacy-exploited Stealthy Model Poisoning Attacks in Federated Learning}

\author{\IEEEauthorblockN{
Md Tamjid Hossain,
Shafkat Islam,
Shahriar Badsha,
Haoting Shen
    }
    
\IEEEauthorblockA{University of Nevada, Reno, NV, USA}

Email: \{mdtamjidh, shafkat\}@nevada.unr.edu,\{sbadsha, hshen\}@unr.edu
}

\maketitle

\begin{abstract}
Federated learning (FL) has become an emerging machine learning technique lately due to its efficacy in safeguarding the client's confidential information. Nevertheless, despite the inherent and additional privacy-preserving mechanisms (e.g., differential privacy, secure multi-party computation, etc.), the FL models are still vulnerable to various privacy-violating and security-compromising attacks (e.g., data or model poisoning) due to their numerous attack vectors which in turn, make the models either ineffective or sub-optimal. Existing adversarial models focusing on untargeted model poisoning attacks are not enough stealthy and persistent at the same time because of their conflicting nature (large scale attacks are easier to detect and vice versa) and thus, remain an unsolved research problem in this adversarial learning paradigm. Considering this, in this paper, we analyze this adversarial learning process in an FL setting and show that a stealthy and persistent model poisoning attack can be conducted exploiting the differential noise. More specifically, we develop an unprecedented DP-exploited stealthy model poisoning ({\tt DeSMP}) attack for FL models. Our empirical analysis on both the classification and regression tasks using two popular datasets reflects the effectiveness of the proposed {\tt DeSMP} attack. Moreover, we develop a novel reinforcement learning (RL)-based defense strategy against such model poisoning attacks which can intelligently and dynamically select the privacy level of the FL models to minimize the {\tt DeSMP} attack surface and facilitate the attack detection. 
\end{abstract}
\vspace{-2pt}
\section{Introduction}
\label{Intro}
\noindent{Federated} Learning (FL), also known as collaborative learning, has caught a lot of attention from the research community since it has been first introduced back in 2016 by McMahan et al. \cite{mcmahan2017communication}. It is mostly because of the inherent privacy protection that FL offers to its users. In the FL process, a model is trained on a diffuse network of edge nodes using their local data; rather than the traditional centralized training fashion. 
This provides a level of data privacy assurance to the users since the confidential data do not leave the edge nodes.

However, the process of FL can be vulnerable to differential attacks (e.g., membership inference attacks (MIA)) which aim to reveal the sensitive information of a node by analyzing the distributed model parameters \cite{geyer2017differentially} or gradients \cite{zhu2020deep}. 
To alleviate this privacy issue, extensive research have been carried out lately, focusing on developing secure multi-party computation (SMPC) \cite{li2020privacy}, trusted execution environments (TEEs) \cite{mo2019efficient}, cryptographic encryption \cite{zhang2020batchcrypt,sadique2021cybersecurity,sadique2019system}, and differential privacy (DP)-based privacy-preservation techniques \cite{dwork2006calibrating, kairouz2019advances, bhagoji2019analyzing} for FL. Among these, DP is considered a very promising technique to preserve the data privacy and prevent MIA \cite{shokri2017membership}. Existing works along this research line include DP-based distributed SGD \cite{abadi2016deep}, local DP (LDP) \cite{wang2019local}, 

Although DP is providing a level of privacy guarantee, an adversary can exploit the DP noise to inject false data into the original data and hide the attack identity exploiting the noise range \cite{giraldo2020adversarial}. In this paper, we investigate this vulnerability of DP-based applications and show that in a differentially private FL setting (we call it `{\tt DPFL}'), a malicious actor can inject the false data either into the differentially private training data (i.e., data poisoning attack \cite{biggio2012poisoning} 
or into the model parameters (i.e., model poisoning attack \cite{bagdasaryan2020backdoor,bhagoji2019analyzing}). More specifically, we demonstrate a stealthy model poisoning attack in the FL model exploiting the noise of the DP mechanism that (1) reduces the overall accuracy of the global federated model, and (2) deceives the traditional anomaly detection mechanisms by hiding the false data into the DP-noise. The results in this paper reveal a new backdoor for stealthy and untargeted model poisoning attacks in FL through the exploitation of the DP mechanism.

\subsection{Motivations}
\label{motive}
\noindent Poisoning attacks in any machine learning (ML) setting can be broadly divided into two major categories: \textit{targeted} and \textit{untargeted} attacks \cite{kairouz2019advances}. Targeted poisoning attacks \cite{bagdasaryan2020backdoor,bhagoji2019analyzing} aim to change the outcome or behavior of the model on particular inputs while maintaining a good overall accuracy on all other inputs, thus makes the attack and defense processes more difficult. On the contrary,
the untargeted model poisoning attacks 
\cite{biggio2012poisoning,fang2020local} have the power to make a model unusable and eventually leads to a denial-of-service attack \cite{fang2020local}. For instance, an adversary may perform untargeted attacks on its competitor's FL model with an intention to make the model unfeasible. 

However, traditional untargeted poisoning attacks mainly utilize the hyperparameters of the targeted model to scale up the effectiveness of the malicious model \cite{bhagoji2019analyzing}. To attain the goal of poisoning, the adversary may use explicit boosting that deforms the weights' distribution, however, then it can be easily detected by the server through simple server-side model checking \cite{pillutla2019robust}. Hence, untargeted model poisoning attacks in a stealthy manner remain an open problem in FL \cite{zhou2021deep}. Moreover, since an FL system usually consists of a huge number of clients and only a portion of clients are chosen for any particular round \cite{sun2019can}, the odds of impacting the global model accuracy significantly by a single malicious contribution is very low. This leads us to the question- \textit{``How can the adversary perform an untargeted model poisoning attacks in a stealthy but persistent fashion?"}.  
Motivated by this, in this paper, we investigate the DP mechanism as a tool to conduct such adversarial poisoning attacks in FL. In the rest of the paper, the `false data injection (FDI)' attack and `model poisoning' attack is mentioned interchangeably.

\subsection{Contributions}
\label{contrib}
In this paper, we show that the DP mechanism is creating a new attack avenue for stealthy false data injection (FDI) or model poisoning attacks in a {\tt DPFL} environment. We name this attack model as `DP-exploited stealthy model poisoning' (in short, {\tt DeSMP}) attacks. Particularly, we make the following contributions:
\begin{itemize}
    \item We demonstrate that DP, as a privacy-preserving tool, is opening a new backdoor for untargeted model poisoning attacks in the FL setting. Our proposed attack strategy ({\tt DeSMP}) is stealthy and persistent in nature. 
    \item To tackle the proposed {\tt DeSMP} attack, we develop a reinforcement learning (RL)-based defense strategy. The proposed RL-based defense approach intelligently selects the differential privacy level for the clients' model update. It also minimizes the attack vectors and facilitates attack disclosure. 
\end{itemize}
Section \ref{preliminariesAndLitRev} of this paper covers preliminaries of FL and a brief review of the related works while section \ref{ProbFormulation} outlines the research problem and threat model. Section \ref{systemModel} formulates the proposed {\tt DeSMP} attack and defense model and their working principle. In section \ref{expAnalysis}, we analyze and evaluate the effectiveness of our proposed model. Finally, in section \ref{conclusionAndFuture}, we conclude the paper with some future research directions.

\section{Preliminaries and Literature Review}
\label{preliminariesAndLitRev}
\noindent{Here}, we discuss the basic mechanism of FL while pointing out some significant contrasting contributions between this work and existing notable research work in adversarial FL. Table \ref{symbol} describes the major symbols used in this paper.
\subsection{Mechanism of Federated Learning with DP}
\label{MechanismDPFL}
\noindent{FL} introduces a collaborative zone for training a model among a set of workers. 
Here, each participating node maintains a local model for its local training dataset.
Additionally, FL incorporates a server that aggregates all the local models to form a global model \cite{mcmahan2017communication}. Furthermore, to tackle MIA through analyzing the model weights, the FL server generally includes a privacy-preserving mechanism such as DP \cite{geyer2017differentially}. Here, DP adds the random Laplacian ($LAP(\frac{\Delta f}{\varepsilon})$) or Gaussian noise ($\mathcal{N}(\theta=0, \sigma^2=\frac{2ln(1.25/\delta).(\Delta f)^2}{\varepsilon^2})$) to the model weights.
\begin{table}[t!]
\centering
\caption{List of major symbols and their description}
\label{symbol}
 \begin{adjustbox}{max width=\linewidth}
\begin{tabular}{||c l c l||} 
 \hline
 Symbols & Description & Symbols & Description \\ [0.5ex]
 \hline\hline
$\tau$ & Accuracy or loss threshold & $\theta$ & Mean\\
$\mu_a$ & Attack impact & $\mathcal{D_M}$ & Measurement data \\ 
$\gamma$ & Attacker's tolerance & $\zeta$ & Norm of model updates \\
$b$ & Batch & $k$ & Participating clients in each round \\
$t$ & Communication round & $f_a$ & PDF of attack distribution \\
$\mathcal{P_{DP}}$ & DP parameters & $f_0$ & PDF of benign Gaussian distribution\\ 
$\mathcal{M}_\mathcal{G}^\mathcal{PS}$ & Final global model & $\mathcal{B}$ & Privacy budget\\ 
$\mathcal{P_{FL}}$ & FL parameters & $\varepsilon$ & Privacy loss \\ 
$\mathcal{M_G}$ & Global model & $\delta$ & Privacy spent in each round \\
$\nabla \mathcal{L}$ & Gradient descent  & $\mathcal{P_{RL}}$ & RL parameters \\
$x$ & Input data & $\Delta f$ or $\mathcal{S}$ & Sensitivity \\
$\mathcal{D_{KL}}$ & Kullback-Leibler divergence & $\sigma$ & Standard deviation \\ 
$\eta$ & Learning rate & $\mathcal{K}$ & Total clients \\
$\mathcal{M_L}$ & Local Model & $w$ & Weights \\ 
$m_l$ & Attacker loss & $f_l$ & Federated loss \\ 
$R$ & Agent reward & $S$ & Agent state \\ 
$a$ & Action & $\alpha$ & Learning rate of RL agent\\
$\pi^*$ & Optimal policy & $\chi$ & Discount factor\\
$Q^*(s,a)$ & converged Q table & $\psi$ & Reward balancing parameter\\[1ex]
 \hline
\end{tabular}
 \end{adjustbox}
 \vspace{-10pt}
\end{table}
Nonetheless, while deploying the DP mechanism, researchers \cite{geyer2017differentially, sun2019can} have suggested using norm clipping or early stopping methods to compensate for the high level of random differential noise and prevent the model to be completely unusable. Once a pre-defined testing criterion (e.g., model accuracy is greater than a threshold or privacy budget exceeds) is met, the server finalizes the global model and stops the training procedure; otherwise, the training process re-initiates.

\subsection{Adversarial Federated Learning}
\noindent{Although} the DP-based FL models do not expose the client's training data to the rest of the world, there exist several attack vectors that an adversary can exploit to perform malicious modification or gain unauthorized access to confidential information. For instance, there could be some malicious clients who might inspect all messages received from the server and then, in the training phases, selectively poison the local models to reduce the efficiency of the global model \cite{kairouz2019advances}. Other examples of the adversarial FL include the targeted and untargeted model poisoning attacks \cite{bhagoji2019analyzing, fang2020local}. However, unlike the centralized ML schemes, the FL systems may employ a large number of untrusted devices which may facilitate the training-time attacks and inference-time attacks \cite{kairouz2019advances}. In this paper, we focus on one of the powerful attack classes which is an untargeted model poisoning attack \cite{fang2020local}. The adversary can conduct this model poisoning attack either by directly manipulating a  client's model or through the widely known man-in-the-middle attack formation leveraging the network and system vulnerabilities \cite{kairouz2019advances}.
\subsection{Related Research Work}
\noindent In this part, we discuss some notable prior research related to the untargeted model poisoning attacks and defenses in FL while outlining some contrasting points with ours.
\subsubsection{Byzantine-robust Aggregation in Adversarial Setting}
Byzantine threat models
\cite{guerraoui2018hidden} 
produce arbitrary outputs for any wrong inputs (either by an honest participant or a malicious actor). These arbitrary outputs can lead to converging the model to a sub-optimal model. Moreover, the Byzantine clients may need to have the white-box access or the non-Byzantine client updates to make their attack stealthy \cite{kairouz2019advances}. Nonetheless, to the best of our knowledge, none of the existing works explore the vulnerabilities of the DP-based applications in tailoring such stealthy attacks. In contrast, we demonstrate that the Byzantine clients or the server can conduct stealthy and persistent untargeted model poisoning attacks by hiding behind the DP mechanism. \textit{In particular, we demonstrate the DP-exploited stealthy model poisoning {\tt (DeSMP)} attacks in an untargeted manner for FL models}.
\subsubsection{DP-assisted FL Frameworks in CPSs}
Another related line of research focus on developing novel FL frameworks for cyber-physical systems (CPSs) such as power IoT \cite{cao2020ifed}, internet of vehicles (IoV) \cite{zhao2020local}, 
smart grids \cite{taik2020electrical} etc. They pave the way for adopting FL into the CPS domain. Particularly, \cite{taik2020electrical} shows that the FL models, coupled with edge computing, perform very efficiently in short-term load forecasting while significantly reducing the networking load compared to a centralized model. \textit{Nevertheless, they do not cover the adversarial analysis of the FL systems for model update poisoning attacks in CPSs}. Since, in CPSs like smart grids, many mission-critical operations depend on the model accuracy, the DP-assisted poisoning attacks may create devastating consequences through the failure of physical layer devices. Therefore, it is non-trivial to investigate the attack surfaces of a {\tt DPFL} model in CPSs. \textit{In this context, we focus on the adversarial analysis of the DP-technique in the CPS domain, which will facilitate the future development of novel and effective defense strategies}.
\subsubsection{Attack Mitigation Strategies in Adversarial DP}
Although some recent works \cite{farokhi2018security, giraldo2020adversarial}  
consider active attacks (e.g., FDI attacks, poisoning attacks, etc.) in DP-based CPSs (e.g., smart grids, transportation systems, etc.), they neither discuss the stealthy model poisoning attacks nor develop any defense strategies based on intelligent decision making for differential privacy level through RL. In particular, they discuss and successively solve the optimal FDI attack problems by developing defense mechanisms based on the anomaly detection schemes for the post-attack phases; instead of taking any initiative to reduce the attack surface beforehand. \hspace{-5pt}

\textit{ In contrast, we analyze the correlation of the DP and FL parameters under adversarial settings; then, leveraging the correlation, we facilitate deployment of the desired level of privacy, utility, and security among the participating nodes in a {\tt DPFL} system through RL. Following the adversarial analysis and our proposed RL-assisted defense strategy, the large-scale poisoning attacks can be detected and the attack surface can be minimized, i.e., the incentive of the attacker can be reduced, which in turn reduces attack motivations while assisting attack prevention.} In short, we develop our RL-assisted defense strategy as a part of the design process (pre-attack phase) to prohibit the untargeted model poisoning attacks. To the best of our knowledge, this is the first work that addresses the DP-exploitation issue in FL setting and successively develops the RL-based defense strategy.

\section{Problem Formulation and Threat Model}
\label{ProbFormulation}
\noindent{Suppose}, we have $\mathcal{K}$ clients, among which $k$ number of clients are selected in each communication round by the server. If the local model updates are $\{\Delta w^1, \Delta w^2,..., \Delta w^i\}$, then the global model update at ($t+1$) communication round is: $\Delta w_{t+1} = \frac{1}{k}\sum_{i=1}^{k}\Delta w^i$ where the $i^{th}$ local model update at ($t+1$) round is: $\Delta w^i_{t+1} = w^i - \Delta w_t$. In an alternative fashion, the loss of the predication can also be calculated as $f_i(w) = \ell(x_i,y_i;w)$ where $(x_i,y_i)$ is the examples set and $w$ represent the weights of global model. Now, according to the {\tt FederatedAveraging} algorithm \cite{mcmahan2017communication}, the objective of the federated server is to minimize the following function: $\min_{w\in\mathbb{R}^d} f(w)\; \text{where}\; f(w) = \frac{1}{k}\sum_{i=1}^{k}f_i(w)$.
The server continues the process until 
the objective is met. 

To introduce DP for preventing model privacy leakage while keeping the model usable, we need to (a) clip the local model updates using the median norm of the unclipped contributions ($\mathcal{S}$) so that the norm is limited and learning is progressing, and (b) add noise from a  DP-preserving randomized mechanism (e.g., Laplace or Gaussian mechanism). Therefore, the new global model update with Gaussian noise at ($t+1$) round becomes: $w_{t+1} = w_t + \frac{1}{k}(\sum_{i=1}^{k} clip(\Delta w_i, \mathcal{S})+\mathcal{N}(0, \sigma^2 \mathcal{S}^2))$. Here, $\sigma^2$ is the variance and $\mathcal{S}$ is the sensitivity of the dataset with respect to the aggregation operation. The value of $\mathcal{S}$ needs to be selected in an optimal way so that the noise variance stays sufficient 
while the aggregated weight's distribution remains as close as possible to the original distribution. Following the related previous research \cite{abadi2016deep, geyer2017differentially}, we set $\mathcal{S}=median\{\Delta w^i\}_{i\in k}$. We draw the noise from a Gaussian distribution with mean ($\theta=0$), variance $\sigma^2$ and PDF (probability density function) as:
\begin{equation}
    f_0(x) = \frac{1}{\sqrt{2\pi}\sigma_x}e^{-\frac{(x-\theta)^2}{2\sigma^2_x}}
    \label{gaussDist}
\end{equation}
However, a malicious actor (if presents) may modify (increase or decrease) the randomized noise in such a fashion that would facilitates \textit{(a) maximum damage}, and \textit{(b) avoid detection}. To perform such stealthy but strong malicious modification, the adversary needs to craft a fake noise profile from either the same or at least, similar distribution function as (\ref{gaussDist}). Earlier research on adversarial differential privacy \cite{giraldo2020adversarial, hossain2021PSU} present us with such optimal attack distribution ($f_a^*$) and impact ($\mu_a^*$) as follows: 
\begin{equation}
    f_a^*(x) = \frac{1}{\sqrt{2\pi}\sigma_x}e^{-\frac{(x-\theta-\sqrt{2\gamma}\sigma_x)^2}{2\sigma^2_x}} \;\text{and}\; \mu_a^* = \theta + \sqrt{2\gamma}\sigma_x
    \label{attackDist}
\end{equation}
Here, a high value of attacker's tolerance ($\gamma$) represents that the adversary does not care to be detected whereas a low value $\gamma$ means the adversary wants to keep a low profile to avoid detection, and thus sacrifices the attack impact ($\mu_a$). More specifically, the adversarial objective of stealthiness can be formulated as: $\mathcal{D}_{\mathcal{KL}}(f_a\left|\right|f_0) \leq \gamma$, where $\mathcal{D}_{\mathcal{KL}}(f_a\left|\right|f_0)$ is the Kullback-Leibler divergence between the PDF of attack distributions ($f_a$) and benign distribution ($f_0$) and indicates the classifier's ability to correctly identify the inputs. Moreover, it can be inferred from (\ref{attackDist}) that during a data or model poisoning attack, the optimal attack impact ($\mu_a^*$) is shifting the benign mean from $\theta$ to $\theta+\sqrt{2\gamma}\sigma_x$. However, $\mu_a^*$ is equal to the actual mean ($\theta$) when $\gamma$ is zero. In short, it implies that when there is no attack or no DP mechanism, the results (in this case, the model weights) remain intact.  The optimal attack distribution of (\ref{attackDist}), $f_a^*$ has been obtained by solving the functional multi-criteria optimization problem of the attacker (i.e., maximum attack while minimum disclosure) and the defender (i.e., maximum privacy with maximum utility). Therefore, if the adversary deviates from the strategy as given by (\ref{attackDist}), he could end up with even lower payoffs \cite{giraldo2020adversarial}.

This observation on adversarial DP analysis motivates us to first raise the question- \textit{``what would be the adversarial impact on the {\tt DPFL} system if the adversary follows the optimal attack strategy, $f_a^*$?"} and then, answer it through theoretical and empirical analysis. Moreover, this potential research problem motivates us to develop a novel and effective defense strategy against such attacks using RL-based intelligent differential privacy level selection. 

\subsection{Threat Model}
\begin{figure}[!t]
    \centerline{\includegraphics[width=\linewidth]{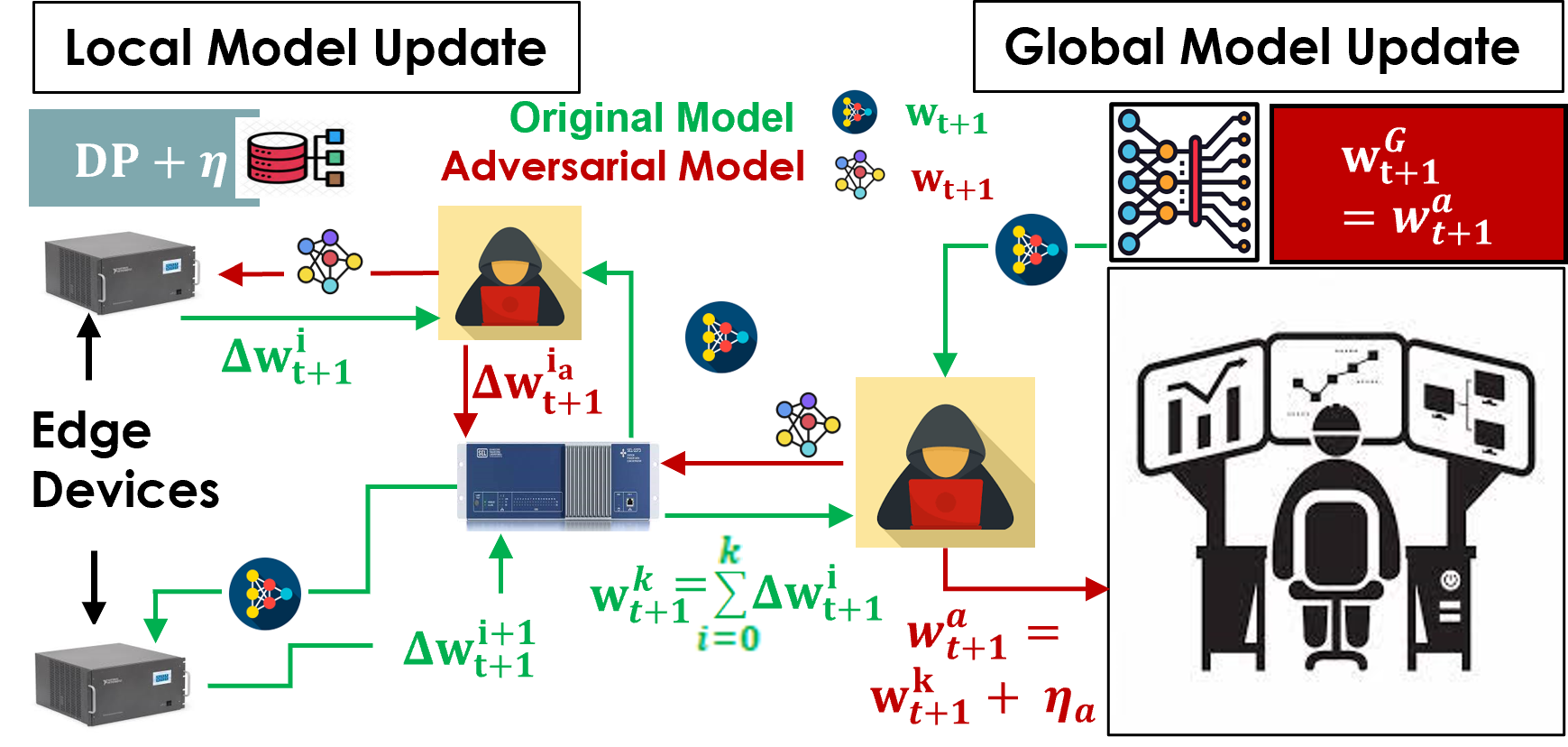}}
    \caption{Threat model: The adversary is exploiting DP to inject false data into the model weights by compromising either the communication path or acting as a server.}
    \label{threatModel}
    \vspace{-15pt}
\end{figure}

\noindent Our proposed threat model has been depicted in Fig. \ref{threatModel}. Here, we are considering a simplified smart grid data transmission architecture which consists of some edge devices (e.g., distribution energy resources (DERs), intelligent electronic devices (IEDs), phasor measurement units (PMUs), etc.), data aggregators (e.g., phasor data concentrators (PDCs)), and a central server. The adversary can mark his presence in- 
(1) the edge nodes (i.e. disguise as an edge device), (2) the communication pathway between the clients and the server, and (3) the server-side. In case of data poisoning attacks, it is convenient for the adversary to compromise some edge devices (i.e., position 1), manipulates local training data, and disguises them as honest edge nodes. However, for model poisoning attacks, the suitable positions for the attacker are positions 2 and 3 since from those positions, the adversary can directly manipulate the FL models through compromising the communication path, sieging the model parameters, and then injecting fake noise into the parameters.

In the proposed setting, we assume that the adversary can manipulate the model updates regardless of the attack vectors (i.e., through man-in-the-middle or server-side attack formation). However, the adversary cannot directly change the models that are already on the server. He has white-box access (i.e., full knowledge of the global and local model parameters). The adversary might have partial knowledge of the training and testing data (i.e., distributions of the data); however, this is not a strict requirement in our threat model. In addition, we assume that the adversary has the knowledge of the imposed DP mechanism and privacy budget ($\varepsilon$). This assumption is particularly important and realistic as many researchers including Dwork et al. \cite{dwork2019differential} emphasize the necessity of publishing the privacy budget in order to increase the trustworthiness of the system.
\section{Modeling {\tt DeSMP} attack and Defense in {\tt DPFL}}
\label{systemModel}
\noindent In this section, we first describe the methodology of our proposed system development from an algorithmic point of view, and then, we model the proposed {\tt DeSMP} attack and RL-assisted defense strategy.
\subsection{Development of {\tt DPFL} Systems}
\noindent{As} discussed in section \ref{MechanismDPFL}, in a {\tt DPFL} system, the global model is first constructed by aggregating all the local models from the randomly selected clients, and then, DP-noise is added into the model parameters to obfuscate the individual contribution of the clients. The working principle of a {\tt DPFL} system with RL-based privacy selection is described through the pseudocodes of algorithm \ref{algo1}. The algorithm simply takes the measured data and the parameters of FL, RL, and DP as input. Then, through some intermediary functions (i.e., $\mathcal{LM}$: local model, $\mathcal{RL}$: reinforcement learning model, $\mathcal{GM}$: global model), the global model is computed. If the computed global model passes the accuracy-test (i.e., accuracy is more than a pre-defined threshold, $\tau$), the global model is finalized and the {\tt DPFL} process completes. 
\begin{algorithm}[!t]
\SetAlgoLined
\textbf{Inputs: }$\mathcal{D}_\mathcal{M}^k, \mathcal{P_{FL}}, \mathcal{P_{DP}}, \mathcal{P_{RL}}$\\
\textbf{Output: }Final Global FL model ($\mathcal{M}_\mathcal{G}^\mathcal{PS}$)\\
  \SetKwFunction{FMain}{$\mathcal{LM}$}
  \SetKwProg{Fn}{Function}{:}{}
  \Fn{\FMain{$\mathcal{D}_\mathcal{M}^k,\mathcal{P_{FL}}$}}{
        $w^k\gets w_t$ $\gets$ $\mathcal{P_{FL}}$\;
\For{local epoch $i = 1,2,...n$}{
    \For{batch $b_i^k \in \mathcal{D}_\mathcal{M}^k$}{
    $w^k\gets w^k-\eta \nabla \mathcal{L}(w^k,b_i^k)$
    }
    }
\KwRet{$\mathcal{M}_\mathcal{L}^k \gets(\Delta w_{t+1}^k,\zeta^k)\gets (w^k -w_t, \lVert \Delta w_{t+1}^k\rVert_2)$}\;
}
 \textbf{End Function}\\

  \SetKwFunction{FMain}{$\mathcal{RL}$}
  \SetKwProg{Fn}{Function}{:}{}
  \Fn{\FMain{$\mathcal{M}_\mathcal{L}^k, \mathcal{P_{RL}}$}}{
        $S_t = (m_l, f_l, \varepsilon_0) \gets(\mathcal{M}_\mathcal{L}^k, \mathcal{P_{RL}})$;\\
        Choose action using epsilon-greedy policy;\\
        Observe $R_{t+1}, S_{t+1}$;\\
        $\pi^{*}(s) = \substack{arg\;max\\\pi}\; Q^{*}(s,a)$;\\
        
\KwRet{$a \gets \pi^{*}(s)$}\;
  }
   \textbf{End Function}\\
  \SetKwFunction{FMain}{$\mathcal{GM}$}
  \SetKwProg{Fn}{Function}{:}{}
  \Fn{\FMain{$\mathcal{M}_\mathcal{L}^k,\varepsilon, \mathcal{P_{DP}}$}}{

$(\Delta w_{t+1}^k,\zeta^k)\gets\mathcal{M}_\mathcal{L}^k$\;
$(\delta, \Delta f, \mathcal{B})\gets\mathcal{P_{DP}},\;\;\; n \gets count(k)$\; 
$\sigma\gets\{\varepsilon,\delta,\Delta f\}$, $\;\;\mathcal{S}^k = median\{\zeta^k\}_{k\in\mathcal{K}}$\;

\lIf{$\delta > \mathcal{B}$}
{
    \Return $w_t$
}
\lElse{$w_{t+1}^k \gets w_t^k + \frac{1}{n}(\sum_{k=1}^{\mathcal{K}} \frac{\Delta w_{t+1}^k}{max(1,\frac{\zeta^k}{\mathcal{S}})} + \mathcal{N}(0, \sigma^2 \mathcal{S}^2)$} 
\KwRet{$\mathcal{M}_\mathcal{G}\gets w_{t+1}^k$}\;
  }
 \textbf{End Function}\\
 $ $\\
     \While{$TestAccuracy(\mathcal{M_G}) < \tau$}{
      \eIf{$\mathcal{D_M}$ is available}{
       $\mathcal{M}_\mathcal{L}^k = \mathcal{LM}(\mathcal{D}_\mathcal{M}^k$, $\mathcal{P_{FL}})$\\
       $\varepsilon = \mathcal{RL}(\mathcal{M}_\mathcal{L}^k, \mathcal{P_{RL}})$\\
       $\mathcal{M_G} = \mathcal{GM}(\mathcal{M}_\mathcal{L}^k, \varepsilon, \mathcal{P_{DP}})$\\
       }{
       wait for $\mathcal{D}_\mathcal{M}^k$ to be available
      }
     }
        \textbf{return} $\mathcal{M}_\mathcal{G}^\mathcal{PS} = \mathcal{M_G}$
\caption{DP- and RL- assisted FL process 
}
 \label{algo1}
\end{algorithm}
\subsubsection{Local Model ($\mathcal{LM}$)}
The $\mathcal{LM}$ function takes the measurement data and learning parameters as inputs. Each client shares a portion of data (i.e., mini-batch) and train the global model with their local data. Finally, the local model and norm updates are calculated and sent back to the server.

\subsubsection{Reinforcement Learning Model $(\mathcal{RL})$}
The purpose of the $\mathcal{RL}$ function is to generate the optimal policy for determining the privacy budget ($\varepsilon$) considering the trade-off among the privacy, utility, and security in a {\tt DPFL} system. The input of this function is the state of the system which comprises of $(m_l, f_l, \varepsilon)$. The function exploits the converged Q-table to determine the optimal action (or value of $\varepsilon$) at each state of the learning process. 

\subsubsection{Global Model ($\mathcal{GM}$)}

The sole purpose of $\mathcal{GM}$ function is to produce the global model ($\mathcal{M_G}$) after each communication round through {\tt FederatedAveraging} procedure \cite{mcmahan2017communication} until the model finally converges around a pre-defined threshold value, $\tau$. The function also checks if the privacy budget is expired on it. Another important task of this function is to clip the gradient to avoid over-fitting or gradient exploding and add Gaussian noise accordingly.





\subsection{Modeling {\tt DeSMP} Attack}
\label{modelDeSMPattack}
\noindent{To} perform the proposed {\tt DeSMP} attack, the adversary needs to choose the level of his stealthiness or attacker's tolerance ($\gamma$). Here, the adversarial goal is to perform the attack so that the model is unusable and ineffective (i.e., converges to a bad-minimum or starts denial-of-service) and the attack is stealthy. For instance, in a classification problem, if the  test inputs are $\{X_i\}_{i=1}^n$, output labels are $\{Y_i\}_{i=1}^n$, global weight vector is $\mathcal{W_G}^t$, global model is $\mathcal{M}_\mathcal{G}^\mathcal{PS}$, benign and attack distributions are $f_0$ and $f_a$ respectively, then the adversarial objective is-
\vspace{-5pt}
\begin{equation}
\label{adversarialGoal}
\begin{split}
\mathcal{A}(\mathcal{W_G}^t) & = \max_{f_a}\sum_{i=1}^{n}[\mathcal{M}_\mathcal{G}^\mathcal{PS}(X_i) - Y_i] \\
& s.t.\;\;\mathcal{D}_{\mathcal{KL}}(f_a\left|\right|f_0) \leq \gamma
\end{split}
\end{equation}
It means the adversary wants to maximize the number of misclassification ($[\mathcal{M}_\mathcal{G}^\mathcal{PS}(X_i) - Y_i]$) while keeping $\mathcal{D}_{\mathcal{KL}}(f_a\left|\right|f_0)$ divergence value below his tolerance level ($\gamma$). 
\begin{figure*}[!b]
    \centerline{\includegraphics[width=\linewidth]{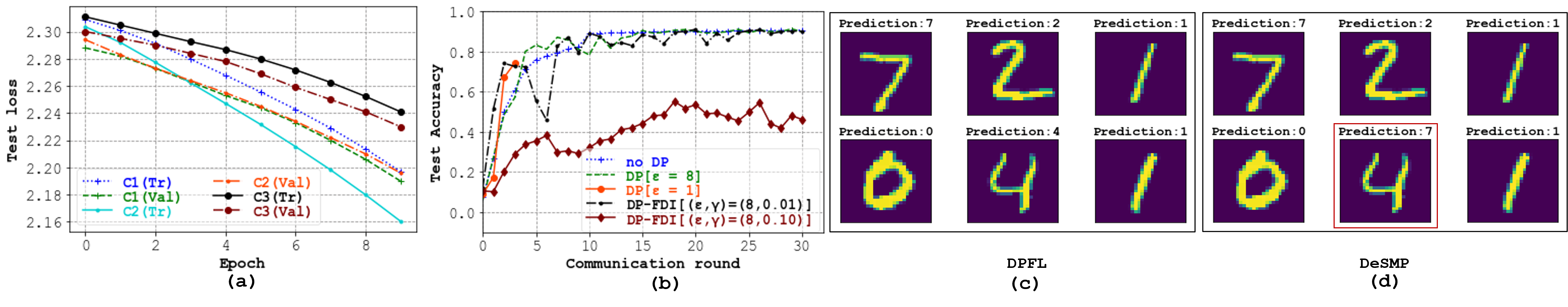}}
    \caption{Evaluation of {\tt DeSMP} attack model on MNIST dataset \cite{deng2012mnist}: (a) training vs validation loss for three random clients (b) test accuracy for non-DP, DP, and FDI-DP data with varying privacy loss ($\varepsilon$) and attacker's tolerance ($\gamma$) (c) {\tt DPFL} model prediction (d) generating incorrect prediction due to {\tt DeSMP} attack}
    \label{MNISTLoss}
\end{figure*}
To achieve this goal, the adversary carefully selects the tolerance value ($\gamma$) and draws noise from the optimal attack distribution, $f_a^*$ as represented by (\ref{attackDist}). In other words, the adversary replaces the benign Gaussian noise mechanism, $\mathcal{N}(\theta,\sigma^2 \mathcal{S}^2)$ by malicious noise adding mechanism $\mathcal{N}_a(\theta+\sqrt{2\gamma}\sigma,\sigma^2 \mathcal{S}^2)$ following (\ref{attackDist}). Here, $\theta$ represents the mean value or location parameter of the Gaussian distribution while $\sigma^2 \mathcal{S}^2$ indicates the scaling factor of the same distribution. By controlling the value of the tolerance level ($\gamma$), the adversary can control the attack impact level ($\mu_a$) and shift the mean value further from the actual value (i.e., $\theta$ to $\theta+\sqrt{2\gamma}\sigma$). In short, increasing/decreasing the value of $\gamma$ increases/decreases the level of noise and vice versa. Nevertheless, since the attack distribution $(f_a^*)$ follows the same statistical properties of a benign distribution $(f_0)$, the adversarial noise $(\mathcal{N}_a)$ as well as the poisonous weights will not be very different statistically from other weights. More specifically, unless the adversary chooses a very large $\gamma$, the proposed {\tt DeSMP} attack will achieve stealthiness while remaining persistent. We empirically observe and evaluate the proposed {\tt DeSMP} attack on the FL models in section \ref{expAnalysis}.

\subsection{Modeling RL-assisted Defense Strategy:}
\noindent{RL}\cite{sutton2018reinforcement} is an adaptive ML algorithm that can facilitate conventional mechanisms with intelligence without the need for any supervision. Distinguishable attributes of RL is a feedback loop (or trial and error) based on the search for optimal action set and delayed rewards. These attributes motivate researchers in deploying RL in divergent sectors, i.e., mmWave communications, smart grid, IoV\cite{9469488}, etc.\\
\indent The addition of DP during the training process will enable the adversary in launching stealthy FDI or poisoning attacks. Moreover, DP will cause degradation in federated accuracy which is difficult to understand and balance the trade-off between privacy, and model performance, both theoretically and empirically\cite{9084352}. On top of this, the FDI attack vector extends the requirement for a trade-off among three different parameters, e.g., privacy, utility, and security. Therefore, selecting the privacy loss ($\varepsilon$) level optimally is a crucial requirement in a {\tt DPFL} system considering the privacy, utility, and security aspects. Our proposed RL-based model assists this optimal privacy policy selection process. Moreover, it defends the learning process from the {\tt DeSMP} attacks by reducing the incentive of the adversary, which in turn reduces attack motivations while assisting attack prevention. In short, in this pa
e., $S = (m_l, f_l, \varepsilon)$.

\indent\textit{Action Space:} We assume that the agent makes a decision in an event-driven manner. By observing the federated environment's current state, the agent makes one of the decisions as described in the action set $(A)$. We can define the action-space as, $A = \{increase,\; decrease,\; static\}$. To fine grain the agent's action making process, we assume that the agent can increase or decrease privacy loss $(\varepsilon)$ by multiple steps (alternatively, a single unit or double unit at any state).

\indent\textit{Reward Function:} 
Reward motivates an agent to make decision towards the learning objectives. For defense against {\tt DeSMP} attack, the objective for the agent is to minimize the maximum attack accuracy as well as maximize the federated accuracy. We assume that the maximum and minimum thresholds are set and regulated by the {\tt DPFL} system designer. We define the reward function for the agent as in equation (\ref{reward}),
\begin{equation}
    \begin{aligned}
    \mathbf{\beta_1}=\psi_1 \frac{f_l^{max}}{f_l} + \psi_2 \frac{m_l^{max}}{m_l} + \psi_3 \frac{1}{\varepsilon}
    \end{aligned}
    \label{reward}
\end{equation}
where $f_l^{max}$ and $m_l^{max}$ denotes the maximum value of FDI attack loss and federated loss whereas $\psi_1$, $\psi_2$, and $\psi_3$ denotes the balancing parameters. \\
\indent Here, we use epsilon-greedy policy\cite{wunder2010classes} for determining the trade-off between exploration and exploitation. We set the initial exploration probability at $1.0$, and gradually reduce the exploration probability over episodes until it matches with the minimum exploration probability (which we assume $0.05$ in this paper).
\section{Experimental Analysis}
\label{expAnalysis} 
\noindent{We} simulate an FL environment in order to test our proposed algorithm. Moreover, for comprehensive evaluation of our proposed {\tt DeSMP} attack model, we focus on the persistence, effectiveness, and stealthiness of the proposed attack under different scenarios for two well-known dataset.

\subsection{Dataset Description and Experimental Setup:}
\noindent{We} utilize the benchmark dataset MNIST (with Non-I.I.D. distributions)\cite{deng2012mnist}, 
Individual household electric power consumption dataset \cite{hebrail2012individual} 
to evaluate our proposed {\tt DeSMP} attack. For MNIST, we have used 10,000 test images to evaluate the performance of the attack model whereas 
In all of the experiments using these two datasets, following the standard FL setup, each selected participants use the SGD (stochastic gradient descent) optimizer to train their local model for internal epoch with local learning rate ($\eta$). 
All of the experiments are done on a server with Intel(R) Core(TM) i7-9700F CPU @ 3.00GHz, 4 NVIDIA GeForce RTX 2060 GPUs with 16 GB RAM each, and Windows 10 (64-bit) OS, with Python 3.8.8 and PyTorch 1.5.1. \vspace{-10pt}
\subsection{Deployment and Evaluation of {\tt DPFL} Model:}
\vspace{-2pt}
\noindent To simulate the {\tt DPFL} environment, we follow some notable prior works \cite{geyer2017differentially, bhagoji2019analyzing,fang2020local} and select the value of some major parameters according to the Table \ref{Parameters}. Moreover, for simplicity, we conduct the experiments with a neural network of three layers. For classification problem (i.e., MNIST), the \textit{Log\_Softmax} activation function has been used on top of the \textit{ReLU} function whereas in regression problems, only \textit{ReLU} function has been used. To add DP-generated noise into the model weights, we modify the {\tt FederatedAveraging} \cite{mcmahan2017communication} procedure according to the Algorithm \ref{algo1}. 
For each experiments, when the privacy budget ($\mathcal{B}$) exceeds, the learning stops and the server finalizes the global model.
\begin{table}[t!]
\centering
\caption{Parameters for FL simulation}
\begin{tabular}{|| l| c| c| c| c| c| c| c ||}
\toprule
\begin{tabular}[c]{@{}l@{}}Parameters/\\ Dataset\end{tabular} & $\mathcal{K}$   & $k$  & $b^i_k$ & $\mathcal{B}$     & $\varepsilon$      & $i$  & $\eta$   \\ \midrule
MNIST                                                         & 100 & 30 & 32  & 0.001 & 0.1-20 & 10 & 0.01 \\
Consumption  & 100 & 30 & 7   & 0.001 & 0.1-20 & 10 & 0.1 \\
\bottomrule
\end{tabular}
\label{Parameters}
\vspace{-10pt}
\end{table}
\begin{figure*}[!b]
    \centerline{\includegraphics[width=\linewidth]{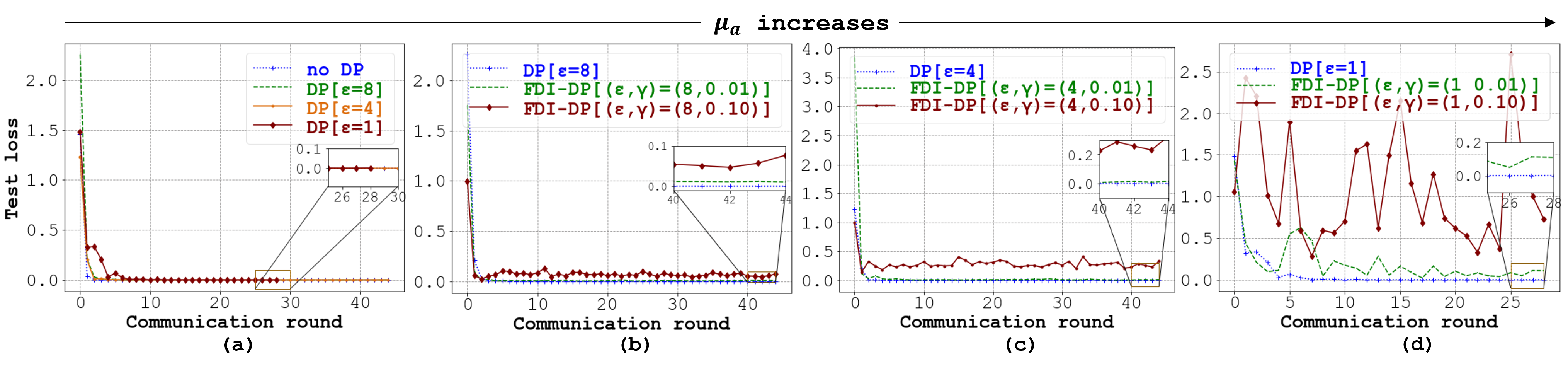}}
    \caption{Evaluation of {\tt DeSMP} attack model on Individual household electric power consumption dataset \cite{hebrail2012individual}: (a) test loss converges even when DP is applied (b) test loss increases as the attacker's tolerance ($\gamma$) increases. (c) more privacy (i.e., small $\varepsilon$) leads to more attack opportunity (d) high privacy and high attacker's tolerance initiates denial-of-service.
    }
    \label{BostonPowerConsumpLoss}
\end{figure*}
 
For MNIST, the training (Tr) and validation (Val) loss of three random clients (C1, C2, and C3) in an arbitrary communication round has been depicted in Fig. \ref{MNISTLoss}(a). It can be inferred that, in each incremental epoch, the training and validation loss is decreasing. Also, from Fig. \ref{MNISTLoss}(b), we can see that the {\tt DPFL} algorithm converges after a few communication rounds. The final accuracy value after round $30$ is around $0.95$. Another important thing to notice is that the privacy budget ($\mathcal{B}$) is spent very quickly if the $\varepsilon$ is small and 
the model can not converge properly. Therefore, it is significantly important to select the privacy loss and budget level (i.e., $\varepsilon$ and $\mathcal{B}$) intelligently and in an optimal way so that the model possesses the desired level of privacy and utility. Likewise, for 
Power consumption dataset, we verify the {\tt DPFL} approach and find similar results. The cost of applying DP (i.e., the `privacy cost`) over the global model loss varies with $\varepsilon$. Thus, more privacy leads to more loss for both the classification and regression problems. 

\subsection{Implementation and Evaluation of {\tt DeSMP} Model:}
\noindent{To} demonstrate the proposed {\tt DeSMP} attack, we replace the benign noise addition mechanism $(\mathcal{N})$ of DP-technique with the adversarial noise addition scheme $(\mathcal{N}_a)$. More specifically, to simulate the behavior of the actual adversary, instead of drawing noise from the benign Gaussian distribution ($f_0$), now we draw noise from attack distribution ($f_a^*$). We can see the impact of such model poisoning action through the DP-FDI curves of Fig. \ref{MNISTLoss}(b). Due to the addition of malicious noise, the overall accuracy has been decreased. However, the degree of model accuracy largely depends on the attacker's tolerance level ($\gamma$). If the attacker chooses to perform more devastating attacks without paying much attention towards achieving the stealthiness, he would select a large $\gamma$ (i.e., $\gamma = 0.10$) and in the process, be able to reduce the accuracy largely. In opposite, selecting a small $\gamma$ (i.e., $\gamma = 0.01$) would give him less payoff in terms of attack impact $(\mu_a)$.

In Fig. \ref{MNISTLoss}(c) and (d), We can further observe the impact of our proposed approach. Fig. \ref{MNISTLoss}(c) reflects the outcome of the proposed {\tt DPFL} model on some randomly selected MNIST image samples whereas \ref{MNISTLoss}(d) depicts the adversarial outcomes through our proposed ({\tt DeSMP}) model for the same samples. Due to the stealthy adversarial noise with $\gamma = 0.10$, only the image of digit `4' has been predicted wrongly as digit `7' while the other digits are predicted correctly. Since we are considering the untargeted model poisoning attack, the adversarial action through the proposed {\tt DeSMP} model may alter the image label differently each time. However, as the overall accuracy does not degrade too much with a low $\gamma$, the malicious action becomes stealthy and goes unnoticed by the anomaly detectors. 

Likewise, the impacts of adversarial action exploiting the DP noise for 
Power consumption dataset have been illustrated in Fig. (\ref{BostonPowerConsumpLoss}). It can be observed that the {\tt DeSMP} attack is also increasing the loss with respect to the increase in $\varepsilon$ and $\gamma$ value. However, for the regression problem, if the raw training data across all the clients are similar and identically distributed, then the attack requires adding more noise (i.e., small $\varepsilon$ and large $\gamma$) in order to achieve the desired level of attack impact. For instance, we can see from Fig. \ref{BostonPowerConsumpLoss}(a) that even after adding the DP mechanism with different $\varepsilon$, the model converges after a sufficient number of communication rounds. It is also desirable since the privacy preserves and the utility remains satisfactory. However, in the presence of an adversary. the loss starts to increase. This phenomena can be observed in Fig. \ref{BostonPowerConsumpLoss}(b)-(d). Moreover, as $\varepsilon$ starts to decrease (i.e., privacy increases) from $8.0$ to $1.0$, the attacker obtains more attack opportunities. From Fig. \ref{BostonPowerConsumpLoss}(b), it can be inferred that the shifting from $(\varepsilon,\gamma) = (8, 0.01)$ to $(\varepsilon,\gamma) = (8, 0.10)$ is increasing the loss by $7$ times (i.e., $0.01$ to $0.07$) whereas in Fig. \ref{BostonPowerConsumpLoss}(c), shifting from $(\varepsilon,\gamma) = (4, 0.01)$ to $(\varepsilon,\gamma) = (4, 0.10)$ is increasing the loss by more than $20$ times (i.e., increasing from $0.01$ to more than $0.20$). Moreover, comparing the red FDI-DP curveS of Fig. \ref{BostonPowerConsumpLoss}(b) and (c), it can be perceived that decreasing $\varepsilon$ by half (from $8.0$ to $4.0$) is increasing the test loss by almost $3$ times ($0.07$ to $0.20$) when tolerance level is relatively high ($\gamma = 0.10$).

Therefore, the attack impact ($\mu_a$) increases significantly with the increment of the attacker's tolerance level, $\gamma$, and the model turns to a sub-optimal model. Eventually, through the {\tt DeSMP} attack, at a very low $\varepsilon$ and high $\gamma$, the {\tt DPFL} model becomes unusable and initiates denial-of-service.
Nevertheless, the proposed {\tt DeSMP} model can also be tailored to conduct more devastating attacks while maintaining stealthiness through hyper-parameter tuning and selectively choosing the FL-parameters that are mentioned in Table \ref{Parameters}.
\subsection{Implementing RL-assisted Privacy Selection}
\begin{figure}[!t]
    \centerline{\includegraphics[width=\linewidth]{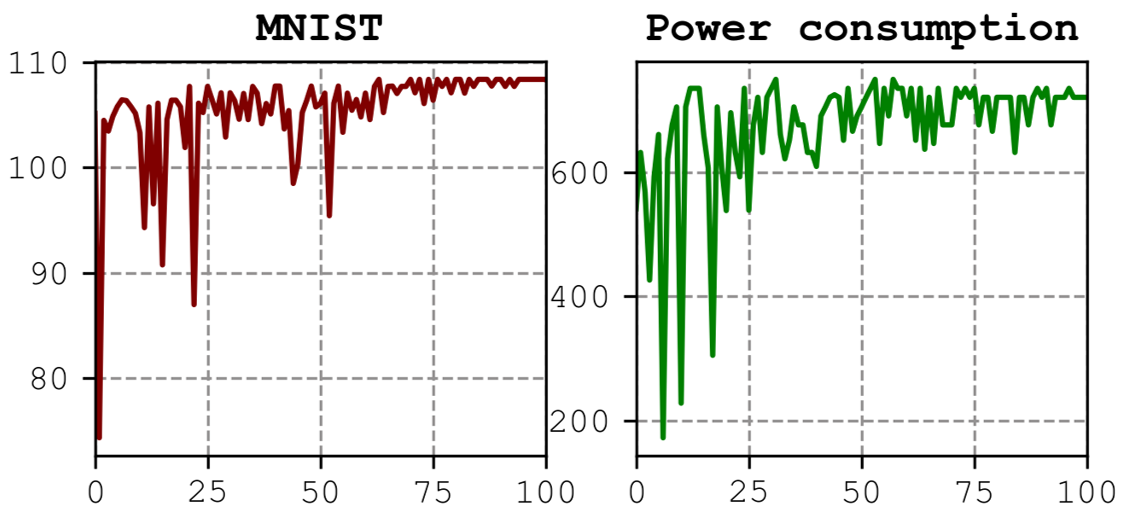}}
    \caption{No. of episodes vs. accumulated rewards for RL assisted privacy selection agent}
    \label{RL}
    \vspace{-10pt}
\end{figure}

\indent Fig. \ref{RL} illustrates the accumulated reward of the defending agent for learning rate $(\alpha = 0.1)$ and discount factor $(\chi = 1)$ for two distinct datasets (MNIST and Power consumption).  The trend in the figure illustrates that the agent learns optimal policy over episodes, and it converges after sufficient episodes are executed. Since we define the reward function such that it takes care of federated loss $(f_l)$, attacker loss $(m_l)$, and privacy loss $(\varepsilon)$, this convergence finds the optimal trade-off policy for the privacy, security, and utility of the system. Since the agent outputs an action (or $\varepsilon$) for each state, we can calculate the standard value of federated loss $(f_l^s)$ for that state. Therefore, if the practical or real-time observed federated loss $(f_l^p)$ differs from the expected (or standard) one, we can infer whether the attack is launched or not. Specifically, if the $f_l^s$ is less than $f_l^p$, we can infer that the large scale  (large $\gamma$) FDI attack is launched; otherwise, the system is not compromised or the degree of FDI attack scale (low $\gamma$) is very low.


\section{Conclusion and Future Works}
\label{conclusionAndFuture}
\noindent{Federated} learning (FL) can be vulnerable to privacy-violating and security-compromising attacks despite having privacy-preserving tools like DP. Model update poisoning is one of such attacks. However, stealthy and persistent model poisoning attacks are difficult to achieve. Motivated by this, in this paper, we analyze the adversarial learning process in an FL setting and show that a stealthy and persistent model poisoning attack can be conducted exploiting the differential noise. More specifically, we develop an unprecedented DP-exploited stealthy model poisoning ({\tt DeSMP}) attack for FL models. Our empirical analysis on both the classification and regression tasks using two popular datasets reflects the effectiveness of the proposed {\tt DeSMP} attack. Moreover, we develop a reinforcement learning (RL)-based novel defense strategy against such poisoning attacks which can intelligently and dynamically select the privacy policy of the FL models to minimize the {\tt DeSMP} attack surface, optimize privacy, security, and utility, and facilitate attack detection. 

In the future, we will extend our defense model for a collaborative multi-agent setting where the team of clients can exploit the learned policy for collaboratively provisioning privacy during the training phase. Although we focus on the untargeted model poisoning attacks in a {\tt DPFL} system in this paper, it would be also interesting to investigate the adversarial impact in targeted model poisoning with our proposed {\tt DeSMP} attack model. We leave it for our future works on adversarial federated learning.
\bibliographystyle{IEEEtran}
\bibliography{bibliography.bib}

\end{document}